\documentstyle[epsfig]{article}

\newcommand{\ba}{\begin{eqnarray}}
\newcommand{\ea}{\end{eqnarray}}
\begin{document}
\pagestyle{plain}

\title{Strange decays from strange resonances}
\author{R. Bijker$^{1}$ and A. Leviatan$^{2}$\\
\mbox{}\\
$^{1}$Instituto de Ciencias Nucleares, 
Universidad Nacional Aut\'onoma de M\'exico, \\
Apartado Postal 70-543, 04510 M\'exico, D.F., M\'exico\\
\mbox{}\\
$^{2}$Racah Institute of Physics, The Hebrew University,
Jerusalem 91904, Israel}

\maketitle

\begin{abstract} 
We discuss the mass spectrum and strong decays of baryon resonances 
belonging to the $N$, $\Delta$, $\Sigma$, $\Lambda$, $\Xi$ 
and $\Omega$ families in a collective string-like model for the 
nucleon. We find good overall agreement with the available data. 
Systematic discrepancies are found for lowlying $S$-wave states, 
in particular in the strong decays of $N(1535)$, $N(1650)$, 
$\Sigma(1750)$, $\Lambda^*(1405)$, $\Lambda(1670)$ and $\Lambda(1800)$. 

\

Se discuten el espectro de masas y los decaimientos fuertes de las 
resonancias bari\'onicas de las familias $N$, $\Delta$, $\Sigma$, 
$\Lambda$, $\Xi$ y $\Omega$ en un modelo colectivo del nucle\'on. 
Se encuentra un buen ajuste con los datos experimentales. Hay una 
desviaci\'on sistem\'atica para los estados en la onda parcial $S$, 
especialmente en los decaimientos fuertes de las resonancias $N(1535)$, 
$N(1650)$, $\Sigma(1750)$, $\Lambda^*(1405)$, $\Lambda(1670)$ y 
$\Lambda(1800)$. 

\

\noindent
PACS number(s): 21.10.Re, 21.60.Ev, 21.60.Fw, 24.60.Lz 
\end{abstract} 

\section{Introduction}

The development of dedicated experimental facilities to probe 
the structure of hadrons in the nonperturbative region of QCD 
with far greater precision than before has stimulated us to 
reexamine hadron spectroscopy in a novel approach in which both 
internal (spin-flavor-color) and space degrees of freedom of 
hadrons are treated algebraically. The new ingredient is the 
introduction of a space symmetry or spectrum generating algebra 
for the radial excitations which for baryons was taken as 
$U(7)$ \cite{BIL1,emff,strong,BIL2}. 
The algebraic approach unifies the harmonic 
oscillator quark model with collective string-like models of baryons. 

We present an analysis of the mass 
spectrum and strong couplings of both nonstrange and strange baryon 
resonances in the framework of a collective string-like $qqq$ model 
in which the radial excitations are treated as rotations and 
vibrations of the strings. The algebraic structure of the model 
enables us to obtain transparent results (mass formula, selection 
rules and decay widths) that can be used to analyze and interpret 
the experimental data, and look for evidence for the existence of 
unconventional ({\it i.e.} non $qqq$) configurations of quarks and 
gluons, such as hybrid quark-gluon states $qqq$-$g$ or multiquark 
meson-baryon bound states $qqq$-$q\overline{q}$. 

\section{Mass spectrum}

We consider baryons to be built of three constituent parts which 
are characterized by both internal and radial (or spatial) degrees 
of freedom. The internal degrees of freedom are described by the 
usual spin-flavor (${\rm sf}$) and color (${\rm c}$) algebras 
$SU_{\rm sf}(6) \otimes SU_{\rm c}(3)$. The radial degrees of 
freedom for the relative motion of the three constituent parts are 
taken as the Jacobi coordinates, which are treated algebraically 
in terms of the spectrum generating algebra of $U(7)$ \cite{BIL1}. 
The full algebraic structure is obtained by combining the radial 
part with the internal spin-flavor-color part 
\ba
{\cal G} \;=\; U(7) \otimes SU_{\rm sf}(6) \otimes SU_{\rm c}(3) ~, 
\ea
in such a way that the total baryon wave function is antisymmetric. 

\begin{figure}
\centerline{\hbox{\epsfig{figure=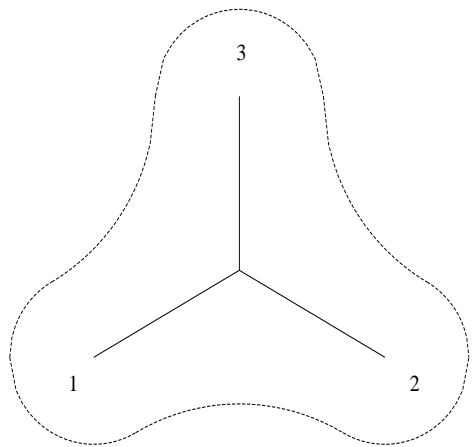,height=0.40\textwidth,width=0.50\textwidth} }}
\caption[]{Collective model of baryons.}
\label{geometry}
\end{figure}

\begin{figure}
\centerline{\hbox{\epsfig{figure=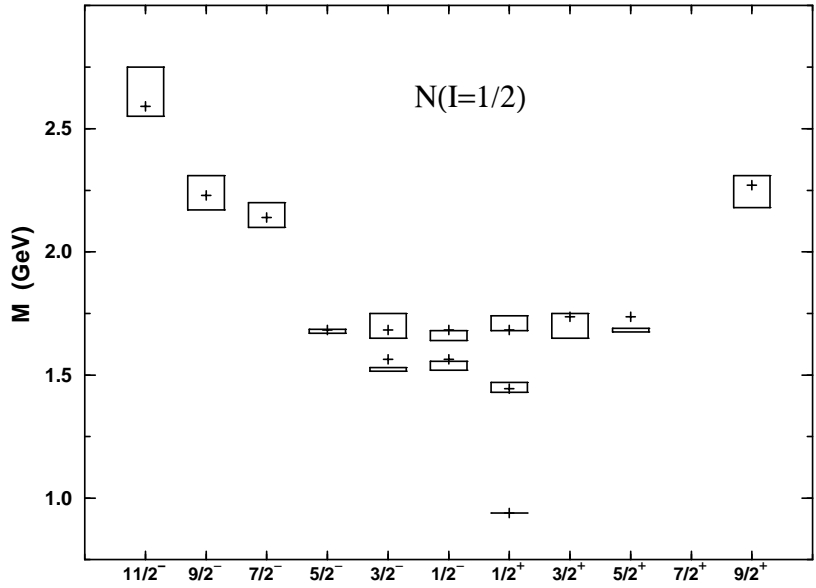,height=0.40\textwidth,width=0.40\textwidth,angle=270} }}
\caption[]{Comparison between the experimental mass spectrum 
of three and four star nucleon resonances (boxes) and the 
calculated masses ($+$). The experimental values are taken from 
\protect\cite{PDG}.} 
\label{nstar}
\end{figure}

\begin{figure}
\centerline{\hbox{\epsfig{figure=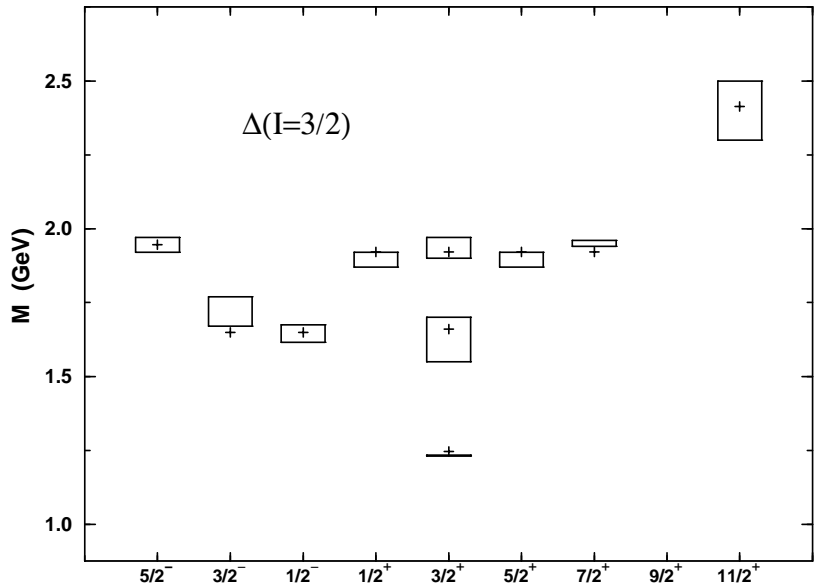,height=0.40\textwidth,width=0.40\textwidth,angle=270} }}
\caption[]{As Fig.~\ref{nstar}, but for $\Delta$ resonances.}
\label{dstar}
\end{figure}

\begin{figure}
\centerline{\hbox{\epsfig{figure=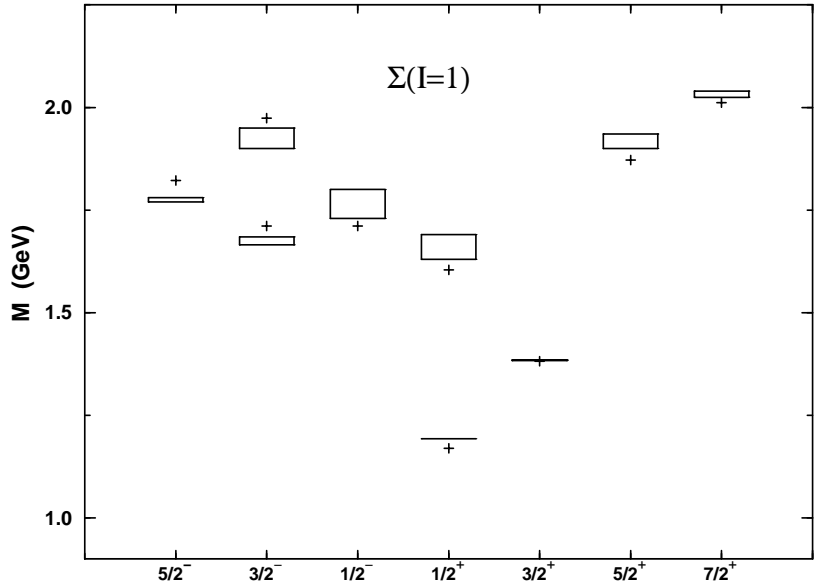,height=0.40\textwidth,width=0.40\textwidth,angle=270} }}
\caption[]{As Fig.~\ref{nstar}, but for $\Sigma$ resonances.}
\label{sstar}
\end{figure}

\begin{figure}
\centerline{\hbox{\epsfig{figure=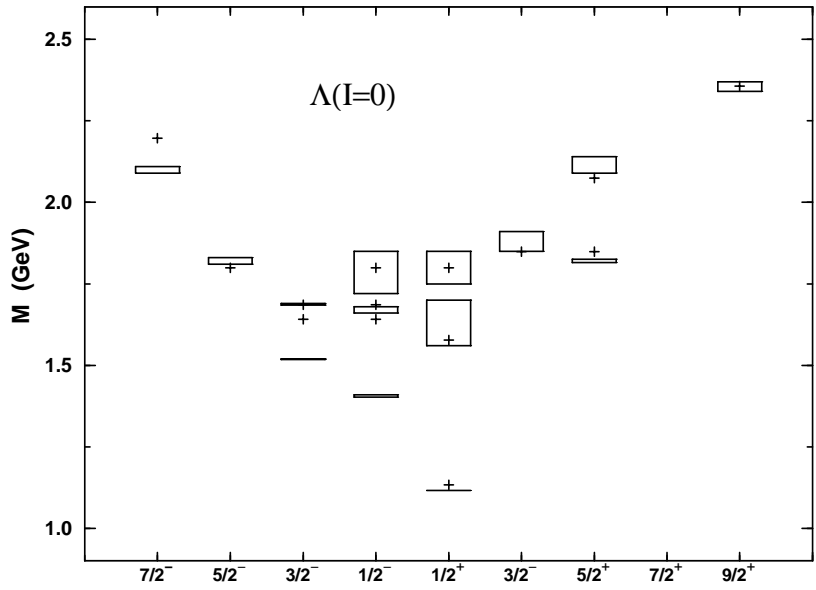,height=0.40\textwidth,width=0.40\textwidth,angle=270} }}
\caption[]{As Fig.~\ref{nstar}, but for $\Lambda$ resonances.}
\label{lstar}
\end{figure}

For the radial part we consider a collective (string-like) model 
of the nucleon in which the radially excited baryons are 
interpreted as rotations and vibrations of the string 
configuration of Fig.~\ref{geometry} \cite{BIL1}. 
The spectrum consists of a series of vibrational excitations 
labeled by $(v_1,v_2)$, and a tower of rotational excitations built  
on top of each vibration. The occurrence of linear Regge 
trajectories suggests to add, in addition to the vibrational 
frequencies $\kappa_1$ and $\kappa_2$, a term linear in $L$. 
The slope of these trajectories is given by $\alpha$. 
For the spin-flavor part of the mass operator we use a 
G\"ursey-Radicati form \cite{GR}.
These considerations lead to a mass formula for 
nonstrange and strange baryons of the form \cite{BIL2}  
\ba
M^2 &=& M^2_0 + \kappa_1 \, v_1 + \kappa_2 \, v_2 + \alpha \, L 
+ a \, \Bigl[ \langle \hat C_2(SU_{ \rm sf}(6)) \rangle - 45 \Bigr]
\nonumber\\
&& + b \, \Bigl[ \langle \hat C_2(SU_{\rm f}(3)) \rangle -  9 \Bigr]
   + c \, \Bigl[ S(S+1) - \frac{3}{4} \Bigr] 
\nonumber\\
&& + d \, \Bigl[ Y - 1 \Bigr] + e \, \Bigl[ Y^2 - 1 \Bigr] 
   + f \, \Bigl[ I(I+1) - \frac{3}{4} \Bigr] ~.
\label{massformula}
\ea
The coefficient $M^2_0$ is determined by the nucleon mass 
$M_{0}^{2}=0.882$ GeV$^2$. The remaining nine coefficients are 
obtained in a simultaneous fit to 48 three and four star 
resonances which have been assigned as octet and decuplet states. 
We find a good overall description of both positive and negative 
baryon resonances of the $N$, $\Delta$, $\Sigma$, $\Lambda$, $\Xi$ 
and $\Omega$ families with an r.m.s. deviation of $\delta=33$ MeV 
\cite{BIL2} to be compared with $\delta=39$ MeV in a previous fit 
to 25 $N$ and $\Delta$ resonances \cite{BIL1}. 
Figures~\ref{nstar}-\ref{lstar} show that there 
is no need for an additional energy shift for the positive 
parity states and another one for the negative parity states, 
as in the relativized quark model \cite{rqm}. 

There are three states which show a deviation of about 100 MeV or 
more from the data: the $\Lambda^*(1405)$, $\Lambda^*(1520)$ 
and $\Lambda^*(2100)$ resonances are overpredicted by 236, 
121 and 97 MeV, respectively. These three resonances are 
assigned as singlet states (and were not included in the fitting 
procedure). An additional energy shift for the singlet states 
(without effecting the masses of the octet and decuplet states) can 
be obtained by adding to the mass formula of Eq.~(\ref{massformula}) 
a term $\Delta M^2$ that only acts on the singlet states. However, 
since $\Lambda^*(1405)$ and $\Lambda^*(1520)$ are spin-orbit partners, 
their mass splitting of 115 MeV cannot be reproduced by such a 
mechanism. In principle, this splitting can be obtained from a 
spin-orbit interaction, but the rest of the baryon spectra shows 
no evidence for such a large spin-orbit coupling. 

A common feature to all $q^3$ quark models is the occurrence of 
missing resonances. In a recent three-channel analysis by the 
Zagreb group evidence was found for the existence of a $P_{11}$ 
nucleon resonance at $1740 \pm 11$ MeV \cite{Zagreb}. The first 
two $P_{11}$ states at $1439 \pm 19$ MeV and $1729 \pm 16$ MeV 
correspond to the $N(1440)$ and $N(1710)$ resonances of the PDG 
\cite{PDG}. It is tempting to assign the extra resonance as one 
of the missing resonances \cite{CLRS}. In the present calculation 
it is associated with the $^{2}8_{1/2}[20,1^+]$ configuration and 
appears at 1713 MeV, compared to 1880 MeV in the relativized 
quark model (RQM) \cite{rqm} (see Table~\ref{missing}). 

A recent analysis of new data on kaon photoproduction \cite{Tran} 
has shown evidence for a $D_{13}$ resonance at 1895 MeV 
\cite{Mart}. In the present calculation, there are several 
possible assignments \cite{BIL2}. 
The lowest state that can be assigned to this new resonance 
is a vibrational excitation $(v_1,v_2)=(0,1)$ with 
$^{2}8_{3/2}[56,1^-]$ at 1847 MeV. This state belongs to 
the same vibrational band as the $N(1710)$ resonance. 
In the relativized quark model 
a $D_{13}$ state has been predicted at 1960 MeV \cite{rqm}. 

\section{Strong couplings}

\begin{figure}
\centerline{\hbox{\epsfig{figure=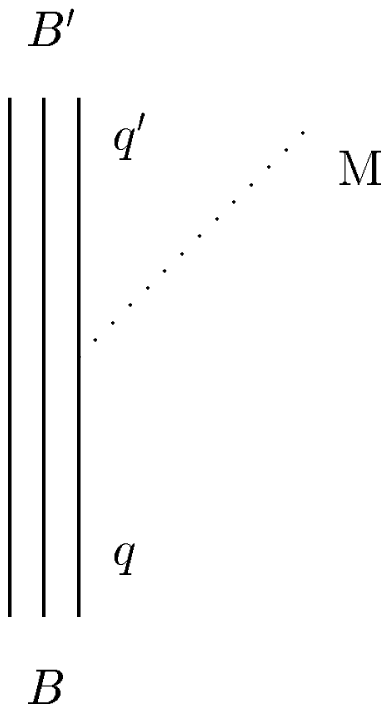,height=\textwidth,width=\textwidth} }}
\vspace{-9cm}
\caption[]{Elementary meson emission}
\label{qqM}
\end{figure}

Decay processes are far more sentitive to details in the baryon 
wave functions than are masses. Here we consider the two-body strong 
decays of baryons by the emission of a pseudoscalar meson
\ba
B \rightarrow B^{\prime} + M ~. 
\ea
We use an elementary emission model in which the meson is emitted 
from a single constituent (see Fig.~\ref{qqM}). The calculation of 
the strong decay widths involves a phase space factor, 
a spin-flavor matrix element which contains the dependence on the 
internal degrees of freedom and a spatial matrix element or form 
factor which in the collective model is obtained by folding with 
a distribution function of charge and magnetization 
\cite{strong,BIL2}. The transition operator that induces the 
strong decay is determined in a fit to the $N \pi$ and $\Delta \pi$ 
channels. It is important to stress that in the 
present analysis the transition operator is the same for {\em all} 
resonances and {\em all} decay channels. The calculations are 
carried out in the rest frame of the decaying resonance. 
For the pseudoscalar $\eta$ mesons we 
introduce a mixing angle $\theta_P=-23^{\circ}$ between the octet 
and singlet mesons \cite{PDG}. 

The calculated decay widths are to a large extent a consequence of 
spin-flavor symmetry and phase space. The spin-flavor symmetry 
gives rise to a selection rule that forbids the decays 
\ba
N(^{4}8[70,L^P]) &\rightarrow& \Lambda(^{2}8[56,0^+]) + K ~, 
\nonumber\\
\Lambda(^{4}8[70,L^P]) &\rightarrow& N(^{2}8[56,0^+]) + \overline{K} ~. 
\label{selrul}
\ea 
This is analogous to the Moorhouse selection rule in 
electromagnetic couplings \cite{Moorhouse}. 
The use of the collective form 
factors introduces a power-law dependence on the 
meson momentum $k$, compared to, for example, an exponential 
dependence for harmonic oscillator form factors. Our results for 
the strong decay widths are in fair overall agreement with the 
available data, and show that the combination of a collective 
string-like $qqq$ model of baryons and a simple 
elementary emission model for the decays can account for the 
main features of the data. As an example, in Table~\ref{del} 
we present the strong decays of three and four star $\Delta$ resonances. 

There are a few exceptions which could indicate evidence for the 
importance of degrees of freedom outside the present $qqq$ model 
of baryons. 

\subsection{Nucleon resonances}

Nonstrange resonances decay predominantly into the $\pi$ 
channel. Phase space factors suppress the $\eta$ and $K$ decays. 
Whereas the $\pi$ decays are in fair agreement with the data, 
the $\eta$ decays of octet baryons show an unusual pattern: 
the $S$-wave states $N(1535)$, $\Sigma(1750)$ and $\Lambda(1670)$ 
all are found experimentally to have a large branching ratio to the 
$\eta$ channel, $74 \pm 39$, $39 \pm 28$ and $9 \pm 5$ MeV, respectively 
\cite{PDG}, whereas the corresponding phase space factor is very small 
\cite{Nefkens}. The small calculated $\eta$ widths ($<0.5$ MeV) 
for these resonances are due to a combination of spin-flavor 
symmetry and the size of the phase space factor. The results of 
our analysis suggest that the observed $\eta$ widths are not due 
to a conventional $qqq$ state, but may rather indicate evidence 
for the presence of a state in the same mass region of a more 
exotic nature, such as a pentaquark configuration 
$qqqq\overline{q}$ or a quasi-molecular $S$-wave resonance 
$qqq$-$q\overline{q}$ just below or above threshold, bound by Van 
der Waals type forces (for example $N\eta$, $\Sigma K$ 
or $\Lambda K$ \cite{Kaiser}). 

The $K$ decays are suppressed with respect to the $\pi$ 
decays because of phase space. In addition, the decay of the 
$N(1650)$, $N(1675)$ and $N(1700)$ resonances into 
$\Lambda K$ is forbidden by the spin-flavor selection of 
Eq.~(\ref{selrul}). For $N(1675)$ and $N(1700)$ 
only an upper limit is known, whereas the $N(1650)$ resonance 
has an observed width of $12 \pm 7$ MeV. However, this resonance 
is just above the $\Lambda K$ threshold which may lead to a 
coupling to a quasi-bound meson-baryon $S$ wave resonance. 

\subsection{Delta resonances}

The strong decay widths of the $\Delta$ resonances are in very 
good agreement with the available experimental data. The same 
holds for the other resonances that have been assigned as 
decuplet baryons: $\Sigma^*(1385)$, $\Sigma^*(2030)$ and 
$\Xi^*(1530)$. For the decuplet baryons there is no $S$ state 
around the threshold of the various decay channels, so therefore 
there cannot be any coupling to quasi-molecular configurations. 
Just as for the nucleon resonances, the $\eta$ and $K$ decays 
are suppressed relative to the $\pi$ decays by phase space factors. 

\subsection{Sigma resonances}

Strange resonances decay predominantly into the $\pi$ and 
$\overline{K}$ channel. Phase space factors suppress the $\eta$ 
and $K$ decays. The main discrepancy is found 
for $\Sigma(1750)$. In the discussion of nucleon resonances 
it was suggested that the $S$ wave state $\Sigma(1750)$ is the 
octet partner of $N(1535)$. It has a large observed $\eta$ width 
despite the fact that there is hardly any phase space available 
for this decay. This may indicate that it has a large 
quasi-molecular component. 

\subsection{Lambda resonances}

Also for $\Lambda$ resonances the $\eta$ and $K$ decays 
are suppressed with respect to the $\pi$ and $\overline{K}$ 
channels because of phase space factors. The strong decays of 
$\Lambda$ resonances show more discrepancies with the data than 
the other families of resonances \cite{BIL2}. 

We have assigned $\Lambda(1670)$ as the octet partner of 
$N(1535)$ and $\Sigma(1750)$. Its decay properties into the 
$\eta$ channel have already been discussed above. 
The calculated $N \overline{K}$ widths of $\Lambda(1800)$, 
$\Lambda(1830)$ and $\Lambda(2110)$ vanish because of the 
selection rule of Eq.~(\ref{selrul}), 
whereas all of them have been observed experimentally \cite{PDG}. 
The $\Lambda(1800)S_{01}$ state has large decay width into 
$N \overline{K}^*(892)$. Since the mass of 
the resonance is just around the threshold of this channel, 
this could indicate a coupling with a quasi-molecular $S$ 
wave. The $N \overline{K}$ width of $\Lambda(1830)$ is 
relatively small ($6 \pm 3$ MeV), and hence in qualitative 
agreement with the selection rule. The situation for the 
the $\Lambda(2110)$ resonance is unclear. 

The $\Lambda^*(1405)$ resonance has a anomalously large decay 
width ($50 \pm 2$ MeV) into $\Sigma \pi$. This feature 
emphasizes the quasi-molecular nature of $\Lambda^*(1405)S_{01}$ 
due to the proximity of the $N \overline{K}$ threshold. 
It has been shown \cite{Arima} that the inclusion of the 
coupling to the $N \overline{K}$ and $\Sigma \pi$ decay 
channels produces a downward shift of the $qqq$ 
state toward or even below the $N \overline{K}$ threshold. 
In a chiral meson-baryon Lagrangian approach with an 
effective coupled-channel potential the $\Lambda^*(1405)$ 
resonance emerges as a quasi-bound state of $N \overline{K}$ 
\cite{Kaiser}. 

\section{Summary and conclusions}

In this contribution we have presented a systematic analysis of 
the masses and the strong couplings of baryon resonances in a 
collective string-like $qqq$ model of the nucleon. 
The algebraic structure of the model, both for the internal 
degrees of freedom of spin-flavor-color and for the spatial 
degrees of freedom, gives rise to transparent results, such as a 
mass formula, selection rules and closed expressions for decay 
widths. 

Wheras the mass spectrum is reasonably well described, the strong 
decay widths are only qualitatively described. The combination of 
a collective string-like $qqq$ model of baryons and a simple 
elementary emission model for the decays can account 
for the main features of the data. 
The main discrepancies are found for the low-lying $S$-wave states,  
specifically $N(1535)$, $N(1650)$, $\Sigma(1750)$, $\Lambda^*(1405)$, 
$\Lambda(1670)$ and $\Lambda(1800)$. All of these resonances 
have masses which are close to the threshold of a meson-baryon 
decay channel, and hence they could mix with a quasi-molecular 
$S$ wave resonance of the form $qqq-q\overline{q}$. 
In contrary, in the spectrum of decuplet baryons there are no 
low-lying $S$ states with masses close to the threshold of a 
particular decay channel, and their spectroscopy is described 
very well. 

The results of this analysis suggest that in future experiments 
particular attention be paid to the resonances mentioned above 
in order to elucidate their structure, and to look for evidence 
of the existence of exotic (non $qqq$) configurations of quarks 
and gluons.  

\section*{Acknowledgements}

This work was supported in part by CONACyT under project 32416-E.

\clearpage

\begin{table}
\centering
\caption[]{Masses of the first three $P_{11}$ states in MeV}
\vspace{15pt}
\label{missing}
\vspace{15pt}
\begin{tabular}{cccc}
\hline
& & & \\
PDG \protect\cite{PDG} & Zagreb \protect\cite{Zagreb} & 
RQM \protect\cite{CLRS} & present \protect\cite{BIL2} \\
& & & \\
\hline
& & & \\
$N(1440)$ & $1439 \pm 19$ & 1540 & 1444 \\
$N(1710)$ & $1729 \pm 16$ & 1770 & 1683 \\
          & $1740 \pm 11$ & 1880 & 1713 \\
& & & \\
\hline
\end{tabular}
\end{table}

\begin{table}
\centering
\caption[]{Strong decay widths of three and four star delta 
resonances in MeV. 
The experimental values are taken from \protect\cite{PDG}. 
Decay channels labeled by -- are below threshold.}
\label{del} 
\vspace{15pt}
\begin{tabular}{lccccc}
\hline
& & & & & \\
Baryon & $N \pi$ & $\Sigma K$ 
& $\Delta \pi$ & $\Delta \eta$ & $\Sigma^* K$ \\
& & & & & \\
\hline
& & & & & \\
$\Delta(1232)P_{33}$ & 116 & -- & -- & -- & -- \\
& $119 \pm 5$ & & & & \\
$\Delta(1600)P_{33}$ & 108 & -- & 25 & -- & -- \\
& $61 \pm 32$ & & $193 \pm 76$ & & \\
$\Delta(1620)S_{31}$ &  16 & -- & 89 & -- & -- \\
& $38 \pm 11$ & & $68 \pm 26$ & & \\
$\Delta(1700)D_{33}$ &  27 & 0 & 144 & -- & -- \\
& $45 \pm 21$ & & $135 \pm 64$ & & \\
$\Delta(1905)F_{35}$ &   9 & 1 &  45 & 1 & 0 \\
& $36 \pm 20$ & & $< 45 \pm 45$ & & \\
$\Delta(1910)P_{31}$ &  42 &   2 &   4 & 0 & 0 \\
& $52 \pm 19$ & & & & \\
$\Delta(1920)P_{33}$ & 22 &   1 &  29 & 1 & 0 \\
& $28 \pm 19$ & & & & \\
$\Delta(1930)D_{35}$ &   0 &   0 &   0 &   0 &   0 \\
& $53 \pm 23$ & & & & \\
$\Delta(1950)F_{37}$ & 45 &   6 &  36 &   2 & 0 \\
& $120 \pm 14$ & & $80 \pm 18$ & & \\
$\Delta(2420)H_{3,11}$ & 12 &  4 &  11 &   2 &   1 \\
& $40 \pm 22$ & & & & \\
& & & & & \\
\hline
\end{tabular}
\end{table}

\clearpage

\clearpage

\vspace{1cm} 

\vspace{3cm}

\clearpage

\vspace{1cm}

\vspace{3cm}

\clearpage 

\begin{thebibliography}{99}

\bibitem{BIL1}
R. Bijker, F. Iachello and A. Leviatan,
Ann. Phys. (N.Y.) {\bf 236} (1994), 69.

\bibitem{emff}
R. Bijker, F. Iachello and A. Leviatan,
Phys. Rev. C {\bf 54} (1996), 1935.

\bibitem{strong}
R. Bijker, F. Iachello and A. Leviatan,
Phys. Rev. D {\bf 55} (1997), 2862.

\bibitem{BIL2}
R. Bijker, F. Iachello and A. Leviatan,
Ann. Phys. (N.Y.) {\bf 284} (2000), 89.

\bibitem{GR}
F. G\"ursey and L.A. Radicati,
Phys. Rev. Lett. {\bf 13} (1964), 173.

\bibitem{rqm}
S. Capstick and N. Isgur, 
Phys. Rev. D {\bf 34} (1986), 2809.

\bibitem{Zagreb}
M. Batini\'c, I. Dadi\'c, I. \v{S}laus, A. \v{S}varc, 
B.M.K. Nefkens and T.-S.H. Lee, 
Phys. Scr. {\bf 58} (1998), 15. 

\bibitem{PDG}
Particle Data Group,
Eur. Phys. J. C {\bf 15} (2000), 1.

\bibitem{CLRS} 
S. Capstick, T.-S.H. Lee, W. Roberts and A. \v{S}varc, 
Phys. Rev. C {\bf 59} (1999), R3002. 

\bibitem{Tran}
M.Q. Tran et al., 
Phys. Lett. B {\bf 445} (1998), 20. 

\bibitem{Mart}
T. Mart and C. Bennhold, 
Phys. Rev. C {\bf 61} (2000), 012201. 

\bibitem{Moorhouse}
R.G. Moorhouse, 
Phys. Rev. Lett. {\bf 16} (1966), 772. 

\bibitem{Nefkens}
B.M.K. Nefkens, 
in `Proceedings of the Fourth CEBAF/INT Workshop: 
$N^*$ Physics', Eds. T.-S.H. Lee and W. Roberts, 
World Scientific, Singapore, 1997, p. 186. 

\bibitem{Kaiser}
N. Kaiser, T. Waas and W. Weise, 
Nucl. Phys. A {\bf 612} (1997), 297.

\bibitem{Arima}
M. Arima, S. Matsui and K. Shimizu, 
Phys. Rev. C {\bf 49} (1994), 2831. 

\end{thebibliography}
\end{document}